\documentclass[onecolumn,noshowpacs,noshowkeys,aps,superscriptaddress,bibliography]{revtex4-1}%
\usepackage{graphicx}
\usepackage{amsmath}
\usepackage{notes2bib}
\usepackage{setspace}

\usepackage{geometry}

\newcommand{\Fig}[1]{Figure \ref{#1}}   
\newcommand{\fig}[1]{fig.\ref{#1}}   

\begin{document}

\title{Time reversal constraint limits unidirectional photon emission in slow-light photonic crystals}

\author{Ben Lang}
\email[Contact me at: ]{bl9453@bristol.ac.uk}
\affiliation{Department of Electrical and Electronic Engineering, University of Bristol, Merchant Venturers Building, Woodland Road, Bristol, BS8 1UB, UK}

\author{Daryl M. Beggs}
\affiliation{Centre for Quantum Photonics, H.H. Wills Physics Laboratory, University of Bristol, Tyndall Avenue, Bristol, BS8 1TL, UK}

\author{Ruth Oulton}
\affiliation{Department of Electrical and Electronic Engineering, University of Bristol, Merchant Venturers Building, Woodland Road, Bristol, BS8 1UB, UK}
\affiliation{Centre for Quantum Photonics, H.H. Wills Physics Laboratory, University of Bristol, Tyndall Avenue, Bristol, BS8 1TL, UK}

\begin{abstract}
Photonic crystal waveguides are known to support C-points - point-like polarisation singularities with local chirality.  Such points can couple with dipole-like emitters to produce highly directional emission, from which spin-photon entanglers can be built.  Much is made of the promise of using slow-light modes to enhance this light-matter coupling.  Here we explore the transition from travelling to standing waves for two different photonic crystal waveguide designs.  We find that time-reversal symmetry and the reciprocal nature of light places constraints on using C-points in the slow-light regime.  We observe two distinctly different mechanisms through which this condition is satisfied in the two waveguides.  In the waveguide designs we consider, a modest group-velocity of $v_g \approx c/10$ is found to be the optimum for slow-light coupling to the C-points.
\end{abstract}


\maketitle

\section{Introduction}

Slow-light photonic crystal waveguides are known to enhance the interaction between optical modes and embedded dipole-like quantum emitters, with recent results showing near-deterministic coupling of $\beta > 98\%$ \cite{Arcari} with a quantum dot.  This enhancement is due to the very tight mode confinement and high density of states in the slow-light waveguide mode \cite{slow_light1, krauss_2007}.  Slow-light is an interference effect that occurs in periodic optical media; at the bandedge (when $\lambda/2$ matches the periodicity of the photonic crystal) the forward and backward travelling components of the wave are equal, and interfere to give a standing wave with group velocity $v_g = 0$.  Just away from the bandedge, these same components are almost equal, and interfere to give an envelope that moves slowly forward with $v_g \ll c$.

Photonic crystal waveguides also support circular-points (or C-points), which are point-like positions in the waveguide where the transverse component of the electric field is equal in magnitude to the longitudinal one, but with a quarter-wave phase-shift (\textit{cf.} polarisation singularities \cite{Nye_cpoints, snom_scan, berry1, nye2}).  Thus it has a circular polarisation \cite{snom_scan}.  Such polarisation structure allows the waveguide to display chirality, despite its physical structure lacking any chiral symmetry \cite{transverse_long, Wheeldon_polarisation}; it is the propagation direction of the light which gives rise to the symmetry breaking required \cite{spin_hall_light}. Note that the ``global chirality'' is zero, even as the ``local chirality'' is non-zero, as the C-points occur in pairs of left-handed and right-handed points.


This local chirality has particular consequences for quantum emitters which tend to be localised, point-like dipole emitters, and can often possess their own chirality. For example, semiconductor quantum dot transitions impart angular momentum on the photons they emit, dependent on the spin state of the electrons and holes within the structure. If the quantum dot is placed at a C-point inside the waveguide, this angular momentum interacts with the chirality to give unidirectional emission dependent on the spin of the electron.  Such spin-dependent unidirectional emission is a very attractive property for quantum information applications, as it allows for the deterministic entanglement of spin and photon path information \cite{young_prl}. Further it can enhance the spontaneous entanglement between systems coupled to the waveguide \cite{entanglement_amplification}.

Directional behavior of this sort has been experimentally demonstrated with quantum dots in photonic structures \cite{Coles_sheffield}, surface plasmons \cite{rodriguez} and atoms coupled to the evanescent fields of optical resonators \cite{atom_switch}, as well as in photonic crystal waveguides \cite{lodahl_chiral, snom_excite}. It is well-known that slow-light enhances the light-matter coupling of emitters to photonic crystals.  Thus, it seems obvious to investigate the use of slow-light to enhance the spin-photon coupling at C-points in photonic crystal waveguides.  However, standing waves (modes with $v_g=0$) are constrained by temporal symmetry and the reciprocal nature of light: that is, the equal forwards and backwards travelling components of the wave destroys propagation direction symmetry breaking, ensuring that any chirality must vanish. This fact places limits on the use of chirality in the slow-light regime.

In this paper, we use calculations of the waveguide eigenmodes to investigate the limit of polarisation chirality in photonic crystal waveguides in the slow-light regime.  We analyse the temporal symmetry constraint in greater depth, showing that the chirality disappears when C-points of opposite handedness approach and annihilate each other as $v_g \rightarrow 0$.  We also present a second photonic crystal waveguide design that lacks the inversion symmetry usually present in waveguides. The crucial feature of this waveguide lies in its photonic bandstructure: there are two waveguide modes which become degenerate at the bandedge. This degeneracy changes the role of the temporal symmetry, allowing both modes to contain chirality, and even C-points at the bandedge.  However, we show that nature conspires against us: despite the existence of chirality in degenerate standing waves, that chirality cannot effectively be used as the waveguide modes are multi-moded near the bandedge.  In other words, C-points of opposite handedness approach in frequency (rather than space) and annihilate at the bandedge.  We find that the optimum group velocity to exploit the C-points is $v_g =c/10$ in the W1 waveguides and $v_g =c/6$ in the glide waveguides.


\section{Waveguide Structures}

\begin{figure}[!h]
\centering\includegraphics[width=5in]{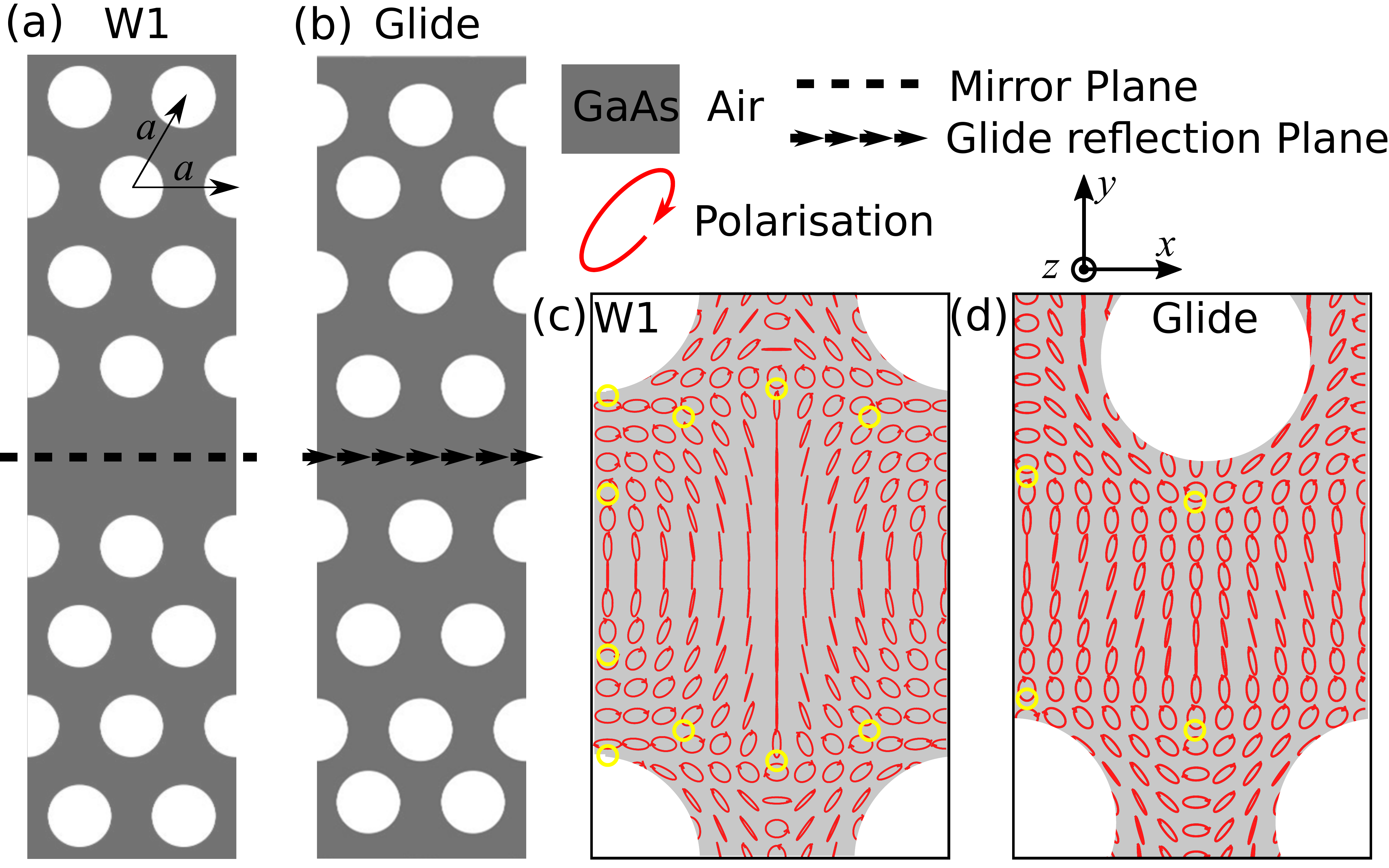}
\caption{Two dimensional refractive-index profiles of the W1 (a) and glide (b) photonic crystal waveguide structures that we consider in this paper. (c, d) The polarisation field (denoted by the polarisation ellipses) of a mode from each waveguide. Yellow circles: C-points or points of circular polarisation}
\label{eps}
\end{figure}

\Fig{eps} shows the two waveguide structures we consider in this paper.  Both waveguides are based on the photonic crystal slab - a two dimensional hexagonal lattice of air holes etched into a thin dielectric membrane.  The lattice constant $a$ describes the separation of the holes.  In this paper we consider waveguides to have a refractive index of $3.46$, a hole radius of $0.3a$ and a membrane thickness of $0.6a$.  This photonic crystal slab has a band-gap between the frequencies of $0.27c/a$ and $0.34c/a$, meaning light at these frequencies is forbidden from propagating through the crystal.  Our photonic crystal slab has values that correspond to using a GaAs membrane at wavelengths around 900nm.

The first type of waveguide we consider is known as a W1 waveguide, so-called as it consists of a single row of missing holes from the hexagonal pattern \cite{w2}, as shown schematically in \fig{eps}\,(a).  The W1 is the archetypical photonic crystal waveguide, and as such its properties have been studied extensively \cite{tfk1, beggs1, noda1, notomi1, mcnab1}.  In particular its dispersion $\omega (k)$ is well known, and consists of a fast light region ($v_g  \lesssim c$) that becomes progressively slower as the bandedge is approached, until at the bandedge there exists a standing wave with $v_g =0$.  The ability to slow light has inspired many applications, as it enhances linear phase shifts, optical nonlinearities and light-matter interactions.  Note that the W1 waveguide is mirror symmetric to inversions about the plane in the centre of the waveguide.

The basic W1 waveguide is also the starting point for the modification of the dispersion relation to produce tailored designs with more desirable optical properties.  For example, changing the size \cite{frandsen} or position \cite{schulz} of some of the holes can significantly reduce the curvature of the dispersion curve, thereby removing group velocity dispersion.  More recently, attempts to engineer the eigenmodes themselves have been made, both to reduce optical loss \cite{ofaolain} and control polarisation properties \cite{lodahl_chiral}.  


The second type of waveguide we will consider is shown in \fig{eps}\,(b).  It is a modified version of the W1, where all the holes on one side of the waveguide are shifted by half a period ($a/2$) along the direction of the waveguide.  This shift removes the inversion symmetry present in the W1.  Instead, the waveguide possesses ``glide'' symmetry, where an inversion about the symmetry plane, followed by a translation along it, leaves the structure unchanged.   The glide-waveguide is further modified by a small shift of the holes closest to the waveguide towards the centre of the waveguide.  This shift is to maximise the extent of chirality present at small group velocities and was introduced after ref. \cite{lodahl_chiral}.

The two photonic crystal waveguides that we consider both have photonic band-gaps and waveguide modes for transverse electric (TE) polarisations: that is, polarisations where the electric field is in the plane of the waveguide.  The nonzero field components are thus ($E_x$, $E_y$, $H_z$) in the $z=0$ symmetry plane of the waveguide.  (More precisely, these are quasi-TE polarisations.  For structures that possess inversion symmetry about the $z=0$ plane, eigenmodes strictly only have $E_z =0$ in the $z=0$ plane of symmetry.  Away from this plane, $E_z$ becomes non-zero.  For details, see \cite{joanopulus_book}).

As periodic structures, photonic crystal waveguide eigenmodes are Bloch modes, and these modes possess strong longitudinal ($E_x$), as well as transverse ($E_y$), components of the electric field profile.  The spatial dependence of both these components, and the phase difference between them, ensures a complex polarisation landscape in the waveguide's vicinity.  The polarisation state at each point can be conveniently summarised by drawing a local polarisation ellipse.  The eccentricity, angle and handedness of the ellipse uniquely describes the relative magnitudes of $E_x$ and $E_y$, as well as the phase between two. \Fig{eps}\,(c) and (d) show the local polarisation ellipses for both the W1 and glide waveguides near the waveguide core.  The points of circular polarisation (C-points) are marked by yellow circles.

\section{Methods}

We calculated the photonic crystal waveguide dispersion $\omega (k)$ and group velocity ($v_g = \frac{d\omega}{dk}$) and corresponding eigenmodes $\mathbf{E}_k (\mathbf{r})$ using a frequency domain eigensolver \cite{mpb}.  In the calculations we used a 2D supercell of dimensions $a$ x $11a\sqrt{3}$ and a grid size of $a/96$, ensuring convergence of the eigenfrequencies to much better than 0.1\% \bibnote{For the glide waveguide slightly modified dimensions of $a$ x $11.3a\sqrt{3}$ were used instead to account for the modified waveguide width and to ensure a correct join between adjacent cells in the $y$-direction.}.  Such a fine calculation grid was chosen in order to locate and track the positions of the C-points with high fidelity.  2D calculations were chosen to save computational resources.  However, we used an effective index approximation \cite{effective_index} that gives a very good agreement with the full 3D calculations; the eigenmodes display a very high ($>95$\%) agreement to those calculated using a full 3D method.  Such a high agreement makes the 2D method suitable for investigating the polarisation properties of the waveguides.

Once the eigenmodes of the waveguide are known, we can analyse their polarisation properties.  Generically, in any 3D monochromatic field, there will exist planes along which the light is linearly polarised (L-planes) and lines along which it is circularly polarised (C-Lines) \cite{Nye_cpoints, Flossmann2008}.  Here, we are concerned with the 2D cross-section through the centre of the waveguide (the $z=0$ plane), meaning that we will find L-lines and C-points where this plane intersects with the L-planes and C-lines. The $E_z$ component of the electric field is zero in this plane of symmetry \cite{joanopulus_book}, which considerably simplifies the polarisation analysis \cite{elipse_in_3d}.  The polarisation properties are quantified by the three Stokes parameters \cite{textbook}: 

\begin{equation}
\begin{aligned}
S_1 (\mathbf{r}) = (|E_x (\mathbf{r})|^2 - |E_y (\mathbf{r})|^2) / S_0 (\mathbf{r}), \\
S_2 (\mathbf{r}) = 2 Re(E_x^* (\mathbf{r}) E_y (\mathbf{r})) / S_0 (\mathbf{r}),\\
S_3 (\mathbf{r}) = 2 Im(E_x^* (\mathbf{r}) E_y (\mathbf{r})) / S_0 (\mathbf{r}).
\end{aligned}
\label{eq1}
\end{equation}

where $\mathbf{r}$ is the position in the $z=0$ plane, $S_0 (\mathbf{r}) = |\mathbf{E}|^2 = |E_x (\mathbf{r})|^2 + |E_y (\mathbf{r})|^2$ is the total electric field strength and $-1 \le S_{1,2,3} \le 1$ describes the position on the Poincar\'{e} sphere.  $S_1$ describes the extent of longitudinal polarisation, such that positions with $S_1 = 1$ have fully longitudinal polarisation ($E_x = S_0$, $E_y = 0$) and positions with $S_1=-1$ have fully transverse polarisation ($E_x = 0$, $E_y = S_0$).  Similarly, $S_2$ describes the extent of linear diagonal polarisation, so $S_2 = 1$ is fully diagonal ($E_x = E_y$), while $S_2=-1$ is fully antidiagonal ($E_x = -E_y$).  $S_3$ describes the extent of circular polarisation, with $S_3 = 1$ being fully right-handed ($E_x = -iE_y$) and $S_3 = -1$ being left-handed ($E_x = iE_y$).  $S_3 (\mathbf{r})$ is of particular interest to us as it also describes the extent of the local chirality.  It is also helpful to identify contours where each Stokes parameter is zero.  Contours of $S_3 = 0$ are known as linear-lines (L-lines), whereas points where $S_3=\pm 1$ are the C-points.  In order to search for the positions of C-points in the waveguide modes, it is useful to note that they occur where the zero contours of $S_1$ and $S_2$ cross \cite{our_PRA, crossing_lines}.

\section{Symmetry condition at the bandedge}

Maxwell's equations are reciprocal - that is they possess time reversal symmetry - in linear, time-invariant media \cite{joanopulus_book, one_way_waveguides, fan_comment}. This symmetry imposes the condition $\mathbf{E}_{\mathbf{k}} (\mathbf{r}) = \mathbf{E}^{*}_{-\mathbf{k}} (\mathbf{r})$ for two counterpropagating modes with wavevectors $\mathbf{k}$ and $-\mathbf{k}$ and group velocities $v_g$ and $-v_g$.


Now consider the time-reversed situations at various points in the polarisation field. It is intuitive to realise that left-handed C-points become right-handed in the counterpropagating mode (and vice versa), whereas transverse/longitudinal points remain transverse/longitudinal, and likewise for diagonal/antidiagonal. Expressed more generally, this means that going from one mode to the counterpropagating one ($\mathbf{k} \rightarrow -\mathbf{k}$):

\begin{align}\label{2}
\begin{split}
\begin{aligned}
&S_1 \xrightarrow{k \to \hspace{0.25pc} -k} S_1\\
&S_2 \xrightarrow{k \to \hspace{0.25pc} -k} S_2\\
&S_3 \xrightarrow{k \to \hspace{0.25pc} -k} - S_3
\end{aligned}
\end{split}
\end{align}

which is easy to verify from equations \ref{eq1} by making the substitution $\mathbf{E}_{\mathbf{k}} (\mathbf{r}) \rightarrow \mathbf{E}^{*}_{-\mathbf{k}} (\mathbf{r})$.

Now, at the bandedge, we have a standing wave with $v_g = 0$, so time-reversal symmetry requires the ``counter-propagating'' mode (also with $v_g = 0$) to be equal to the forward one, $\mathbf{E}_{\mathbf{k}} (\mathbf{r}) = \mathbf{E}^{*}_{-\mathbf{k}} (\mathbf{r}) = \mathbf{E}_{-\mathbf{k}} (\mathbf{r})$ (for non-degenerate modes). An inspection of equation \ref{2} reveals that one consequence of this symmetry is that $S_3 = -S_3 = 0$ for standing waves, and all chiral components of the polarisation must vanish. (This is very similar to the argument that leads to the conclusion that $v_g=-v_g=0$ at the bandedge.) Away from the bandedge, in the slow-light regime, we therefore have a situation where the value of $S_3$ approaches zero everywhere, but how this occurs for both types of waveguide is quite remarkable.

\section{W1 Waveguide}

\begin{figure}[!h]
\centering\includegraphics[width=5in]{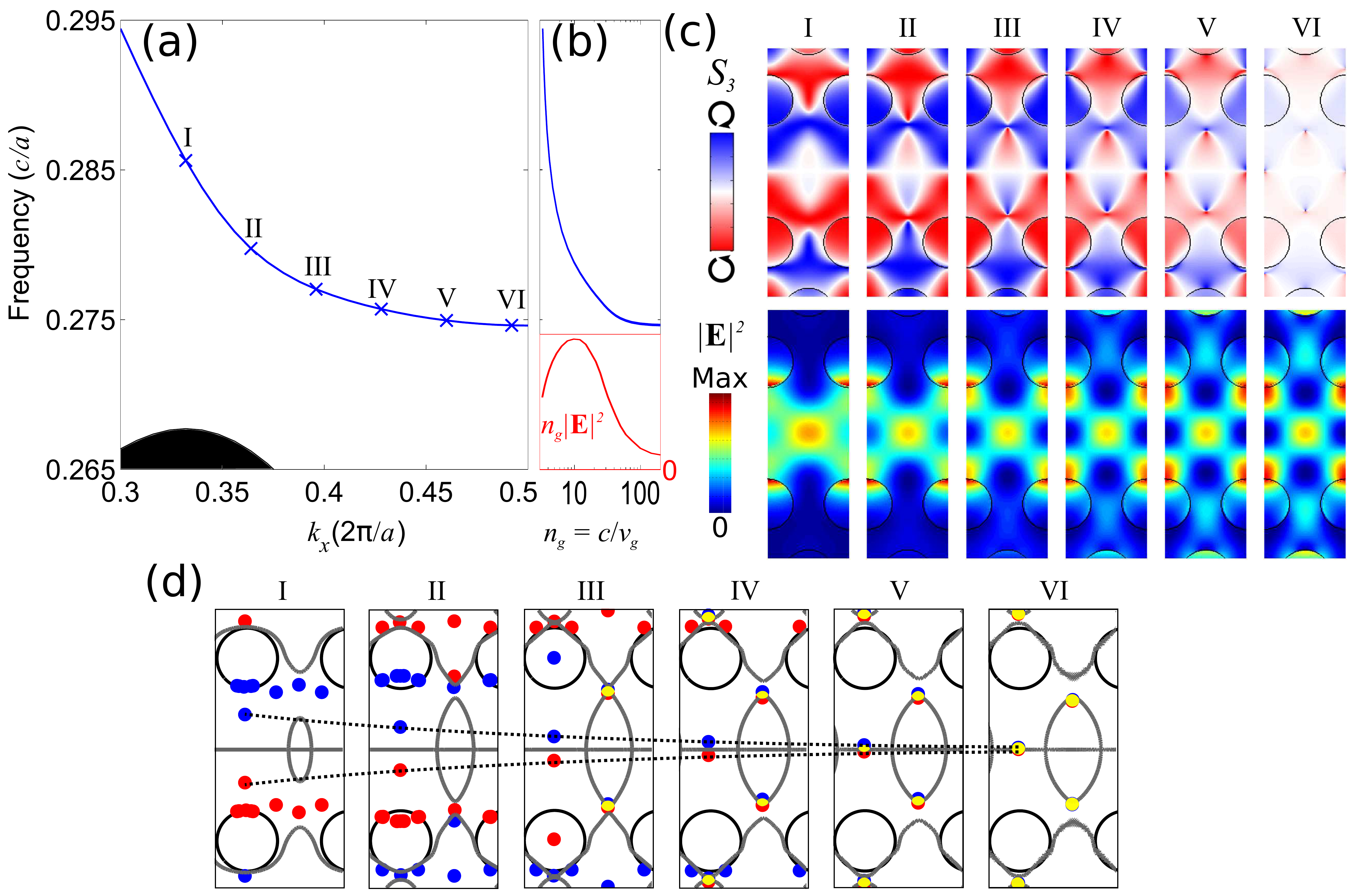}
\caption{The annihilation of C-points at the bandedge in a W1 waveguide. (a) The dispersion of the waveguide mode in the photonic band-gap, and (b) the group index $n_g = c/v_g$ of the fundamental guided mode (note the log scale). The red inset shows the value $|\mathbf{E}|^2 n_g$ at the position of the central C-points as a function of $n_g$.  (c) The electric-field power $|\mathbf{E}|^2$ (bottom row)  and extent of circular polarisation $S_3$ (top row) for the modes I, II, ..., VI indicated in (a). (d) The C-points (red and blue) and L-lines (grey lines) for the same modes. The dashed lines are to guide the eye to the approaching pair of C-points. When two C-points get so close that their markers overlap, the overlap is shaded yellow.}
\label{w1}
\end{figure}

\Fig{w1}\,(a) and (b) show the calculated dispersion $\omega (k)$ and group index $n_g = c/v_g$ for the fundamental waveguide mode of the W1 waveguide.  The group index diverges towards infinity at the bandedge ($k=\pi/a$), indicating that the mode is a standing wave. \Fig{w1}\,(c) shows the maps of $S_3(\mathbf{r})$ and $|\mathbf{E}|^2$ for modes of decreasing group velocity, and we see the general trend that the chirality is washed out as the group velocity decreases. 

However, notice in the final panel, where $v_g=c/300$, we still see points where $S_3 = \pm 1$, indicating the continued existence of C-points.  \Fig{w1}\,(d) shows the positions of the C-points found in the modes marked I, II, ..., VI in (a).  Left-handed C-points are labeled blue ($S_3=-1$), whereas right-handed ones are red ($S_3=+1$).  L-lines ($S_3=0$) are marked by grey contours.  C-points always occur in pairs of opposite handedness, separated by an L-line.  As we approach the bandedge, we observe the C-points of opposite handedness approach each other, until they annihilate at the position of the L-line.  As a guide to the eye, we have marked the approach of the C-points closest to the waveguide centre with a dashed line. (Far from the bandedge we also see C-points with the same handedness come together in groups of three, two of which are eliminated; similar behavior can be seen in ref. \cite{three_to_one,three_to_one2}.)

In fact, C-points still exist for any wavevector arbitrarily close to the bandedge; it is only at the bandedge that they collide and disappear.  However, as the bandedge is approached, the left- and right-handed C-points approach each other until they are very close (the pixelation in our calculations means they are poorly resolved for $v_g < c/300$).  Note also that when two C-points of opposite handedness annihilate on an L-line, a point at which the electric field strength is zero results, referred to as a V-point \cite{Schoonover2006} or $\Sigma$-point \cite{three_to_one2}. Thus, as the two C-points approach the total electric field strength $|\mathbf{E}|^2$  becomes very small in their vicinity, as can be seen in the inset to \fig{w1}\,(b).  The total coupling strength of a dipole emitter placed at a C-point is proportional to the local density of states (LDOS), which is proportional to the product $n_g |\mathbf{E}|^2$, where the electric field strength is evaluated at the position of the C-point.  As we approach the bandedge, the fall off of the electric field strength $|\mathbf{E}|^2$ at the C-point is faster than the growth in $n_g$, and so the LDOS falls.  We plot the product $n_g |\mathbf{E}|^2$ in the inset to \fig{w1}\,(b), and it is clear that there is an optimum point at which to exploit the coupling of an emitter to the waveguide mode at the C-point.  In the W1 design considered here, the optimum occurs for modes with a group velocity of $v_g = c/10$.  Although this figure is no doubt dependent on the design parameters of the waveguides (and could therefore be optimised to slower group velocities and higher LDOS), the point remains that there is an optimum at finite values of $n_g$.  Therefore there is little advantage to using the C-points in the slow-light regime as the key advantage provided by the slow-light (increased LDOS and high coupling strength) is counteracted by a concomitant decrease in local field intensity caused by the approach of oppositely handed C-points.


At the bandedge itself all the C-points are annihilated and the polarisation ellipse at each point contracts to some orientation of linear polarisation. This standing wave is it's own time reverse.

\section{Glide-symmetry waveguides}

\begin{figure}[!h]
\centering\includegraphics[width=5in]{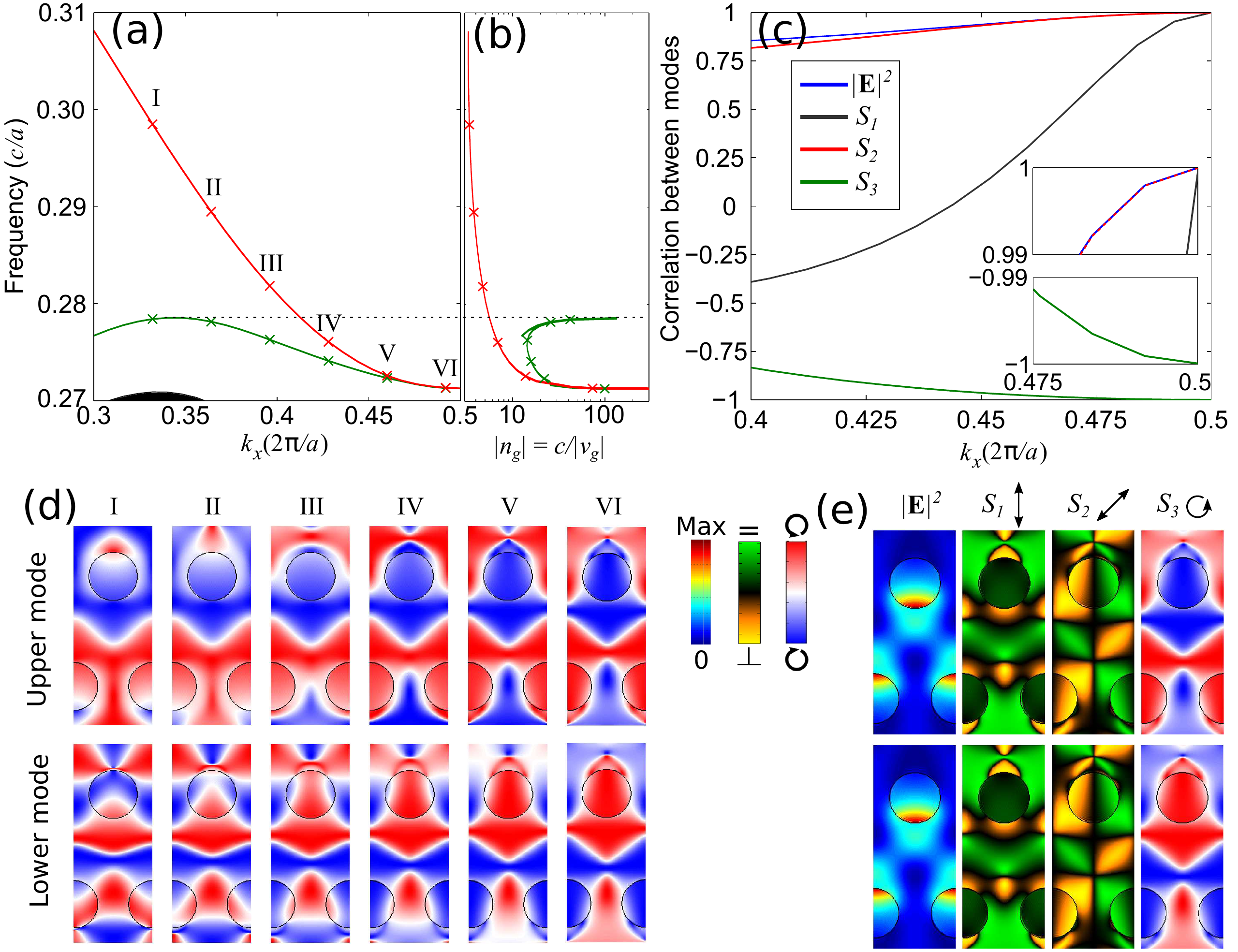}
\caption{The bandedge of the glide symmetry waveguide. (a) The calculated dispersion $\omega(k)$ and (b) the group index $n_g = c/v_g$ of the two waveguide modes. (c) The correlation between the two modes in power and each of the Stokes parameters. Insets: zoomed in plots of the bandedge region. (d) The map of $S_3$ or chirality for  the modes I, II, ..., VI indicated in (a) for the lower and upper frequency mode. Both retain significant chirality far into the slow-light region, but become increasingly anti-correlated. (e) The full electric field profile of the two degenerate modes at the bandedge.}
\label{glide}
\end{figure}

We now explore the bandedge transition in the glide waveguide structure.  \Fig{glide}\,(a) shows the calculated bandstructure of the glide waveguide.  The fundamental mode of the former W1 waveguide has now become two modes that are degenerate at the bandedge, as required by the glide symmetry. (The glide symmetry operation belongs to a class of nonsymmorphic space groups that ensures that any mode must be two-fold degenerate at the bandedge \cite{Adam_Mock}.) As the glide waveguide no longer possesses inversion symmetry, the two modes are neither even nor odd, but instead their symmetry can be classified by decomposing their spatial Fourier components.  The lower mode has spatial Fourier components that alternate even/odd/even/odd/..., while the upper mode's alternate odd/even/odd/even/... \cite{Adam_Mock}.  \Fig{glide}\,(b) shows the group index $n_g = c/v_g$ for the two modes as a function of the frequency.  From \fig{glide}\,(a) and (b), note that the two waveguide modes are indeed degenerate at the bandedge, but also that the frequencies of the two modes overlap for a significant portion of the slow-light regime: i.e. the waveguide becomes multi-mode near the bandedge.

As these two modes approach each other in frequency, they also become more similar, as can be seen in \fig{glide}\,(d), until at the bandedge one should be the time-reversal of the other so that $\mathbf{E}_{\mathbf{k=\pi/a},lower}=\mathbf{E}^*_{\mathbf{k=\pi/a},upper}$.  In \fig{glide}\,(b), we explore the correlation between the two modes in the vicinity of the bandedge.  We quantify the correlation with $-1 \le C \le 1$ which has a value of $+1$ for two fields that are identical (to within a positive factor), and $-1$ when the two fields are identical negatives of each other.  A value of $C=0$ indicates no correlation \bibnote{If a particular Stokes parameter take the values $S_A(\mathbf{r})$ in mode A and $S_B(\mathbf{r})$ in mode B, then the two-dimensional correlation between them is given by $C = \frac{\sum_{\mathbf{r}} (S_A(\mathbf{r}) - \bar{S}_A) (S_B(\mathbf{r}) - \bar{S}_B) }{ \sqrt{ (\sum_{\mathbf{r}} (S_A(\mathbf{r}) - \bar{S}_A)^2) (\sum_{\mathbf{r}} (S_B(\mathbf{r}) - \bar{S}_B)^2) } } $, where $\bar{S}_{A,B}$ is the average of $S_{A,B}(\mathbf{r})$ over $\mathbf{r}$. In our case, the mode A and B refer to the upper and lower waveguide modes.}.

On approach to the bandedge the two modes converge in frequency and become more correlated. In particular, the correlation $C$ for $S_{0,1,2}$ approach 1, whereas those for $S_3$ approach $-1$. This is the signature that one mode is the time reverse of the other (\textit{cf.} equation \ref{2}). At the bandedge itself, the lower and upper modes have the same frequency and are each other's time-reversals.  In particular, note that having degenerate modes lifts the condition that $S_3=0$ and allows the existence of chirality, even in the standing wave with $v_g=0$.  Observing the chirality of these two modes is instructive; in the regions where one is left-handed the other is right-handed and vice-versa (\fig{glide}\,(d)) \bibnote{This transition is complicated by the presence of an (anti)-crossing between the two guided modes at $ka/2\pi\approx 0.476$, before they come back together for a degeneracy at $ka/2\pi=0.5$. It is also made more complicated by the fact that at the bandedge itself any linear combination of modes is itself a mode. The modes found by the eigensolver were not in fact a time-reversal pair. However we found such a pair by taking linear complex combinations of the eignesolver's solutions.}.

Importantly though, using the C-points present in the glide waveguide at $v_g = 0$ will not result in unidirectional emission dependent on spin.  Nature conspires against the use of the C-points present at the bandedge: whereas in the W1 waveguide, the left- and right-handed C-points approached each other in position and annihilated at the bandedge, in the glide waveguide they approach in frequency and annihilate at the bandedge.  This unfortunate fact is caused by the multi-mode nature of the waveguide at the bandedge.  The two degenerate modes have oppositely handed C-points at identical locations and frequencies.  For example, at the location of a right-handed C-point in the lower mode, there will be a left-handed C-point in the upper mode with the same frequency and group velocity.  Thus a right-handed circular dipole (such as a spin-down electron transitioning to the ground state in a quantum dot) placed at this location will only emit light into the forwards direction of the upper mode.  However, it will also emit an equal power into the backwards direction of the lower mode, resulting in no overall directionality.

\begin{figure}[!h]
\centering\includegraphics[width=3.9in]{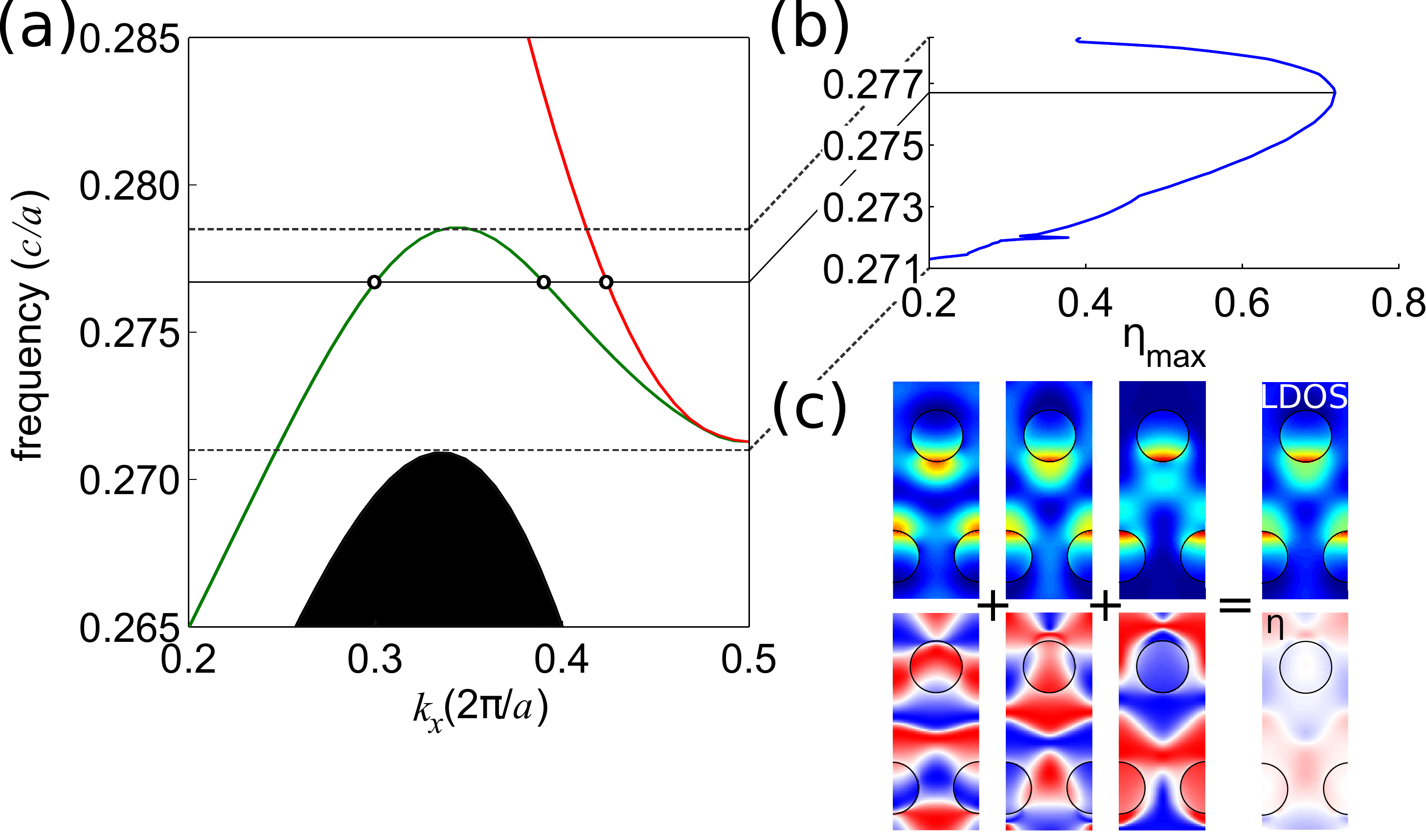}
\caption{The effects of coupling to multiple modes on directionality. (a) The band diagram of the glide waveguide.  (b) The maximum value of $\eta$ in the vicinity of the waveguide as a function of frequency.  (c) $|\mathbf{E}|^2$ and $S_3$ for the three modes at the same frequency in this waveguide (indicated in (a) by yellow markers). $S_3$ in the first panel has been inverted to account for the fact the group velocity of the mode is in the opposite direction to the other two. The final panel is the resulting mixture of these modes.}
\label{multimode}
\end{figure}

Before proceeding further, it will be useful to introduce the directionality, $-1 \le \eta \le +1$, which is defined as the difference between how much power is emitted by a right-handed circular-dipole-like emitter into the forwards and backwards waveguide modes, normalised by the total power emitted into the waveguide.  Values of $\eta = \pm 1$ indicate full emission into one direction or the other (the ideal), whereas $\eta = 0$ indicates no overall directionality.  $\eta$ also gives the degree of spin-path entanglement when used with a quantum emitter. Note that left-handed dipoles reverse the sign of the directionality $\eta$.

Away from the bandedge, in the slow-light regime, there is a frequency band where the dispersion of the lower and upper modes overlap: the waveguide is multi-moded in this region.  \Fig{multimode}\,(a) shows the dispersion of the lower (green line) and upper (red line) modes over a wider frequency range than presented in \fig{glide}\,(a).  The multi-moded region is delimited by the horizontal dashed lines.

In a single-mode waveguide, the directionality $\eta$ is simply given by the value of $S_3$ at the location of the dipole.  In a multi-mode waveguide, however, the emitted power will go into each mode in proportion to its contribution to the local density of states (LDOS).  Thus the multi-mode directionality is calculated by summing the contributions from each mode present at the relevant frequency:

\begin{equation}
\eta (\mathbf{r}) = \frac{\sum_i |\mathbf{E}_{i}(\mathbf{r})|^2 n_{g,i} S_{3,i}(\mathbf{r})}{\sum_i |\mathbf{E}_{i}(\mathbf{r})|^2 |n_{g,i}|},
\label{multimode_eq}
\end{equation}

As can be seen in \fig{multimode}\,(a), in the multi-mode region of interest there are three modes to consider at each frequency: the two near the bandedge from the lower and upper bands, and the third away from the bandedge with an opposite sign of the group velocity.  These three modes are marked with yellow dots for an example frequency ($0.2767c/a$) indicated by the solid horizontal line.  Maps of the electric field intensity $|\mathbf{E}(\mathbf{r})|^2$ and chirality $S_3 (\mathbf{r})$ are shown in \fig{multimode}\,(b), along with the total directionality of the mixture, as calculated from equation \ref{multimode_eq}.  Here, we can clearly appreciate that, although C-points exist in all three modes, the directionality $\eta$ never reaches above 0.72.  \Fig{multimode}\,(c) shows a plot of the peak directionality as a function of the frequency, and for our glide waveguide the directionality peaks at 0.72 in the multi-mode region.  Thus the multi-mode slow-light regime of our glide waveguide design is not suitable for the exploitation of directional effects and spin-path entanglement, and we must move to frequencies where the waveguide is just a single mode, limiting us to group velocities $v_g < c/6$.

\section{Conclusion}

In summary, we have used calculations of the eigenmodes of photonic crystal waveguides to investigate the local chirality effects in the slow-light regime.  Due to the reciprocal nature of light, each mode must have a time-reversal partner, meaning that non-degenerate standing waves at the bandedge must not contain any chirality.  We show that this requirement is met in W1 photonic crystal waveguides when pairs of left- and right-handed C-points collide and annihilate at the bandedge.  C-points exist away from the bandedge, but as they approach each other, their proximity means that the electric field strength is small.  We find that the fall-off in the local density of states due to this decreasing electric field strength is faster than the increase due to the decreasing group velocity, meaning that the local density of states (and therefore the coupling strength to dipole-like emitters) has a maximum value at $v_g \approx c/10$.

We also investigated a photonic crystal waveguide with glide-symmetry and degenerate modes at the bandedge.  This degeneracy allows the two modes to be the time-reverse partners of each other, allowing the existence of chirality and C-points even in standing waves at the bandedge.  However their time-symmetry requires the two degenerate modes to possess oppositely-handed C-points at the same location and same frequency, meaning that neither one can be exploited in isolation for applications.  In fact, as the waveguide is multi-moded in the slow-light regime, we see that the left- and right-handed C-points approach and annihilate in frequency rather than in space.  We note that the lowest group velocity for useful C-points is $v_g \approx c/6$, corresponding to when the waveguide supports a single mode.

Our analysis of C-points in the slow-light regime highlights some of the difficulties to be overcome to exploit chirality in quantum information applications.  Both photonic crystal designs considered here can provide useful functionality only down to very modest slowdown factors of $v_g \approx c/10$.  The mechanisms that cause the annihilation of the C-points in the waveguides considered here are likely to be generally applicable to other photonic crystal and slow-light designs.  However, we believe that alternative designs can be found to extend this useful range down to smaller group velocities, thus enhancing the light-matter interaction with the local chirality.  In fact, it would be interesting to examine how to quantify the trade-off between the increase in LDOS due to smaller group velocities and the directionality given by the local chirality in the waveguide. 

\vspace{2pc}

This work has been funded by the project SPANGL4Q, under FET-Open grant number: FP7-284743. RO was sponsored by the EPSRC under grant no. EP/G004366/1. DMB acknowledges support from a Marie Curie individual fellowship QUIPS. BL gratefully acknowledges funding from EPSRC (DTA). This work was carried out using the computational facilities of the Advanced Computing Research Centre, University of Bristol http://www.bris.ac.uk/acrc/.

\bibliographystyle{unsrt}
\bibliography{bibliogrpahy}

\end{document}